\documentclass{article}
\usepackage{spconf,amsmath,graphicx}
\usepackage{multirow}
\usepackage{cleveref}
\usepackage[table]{xcolor}
\usepackage{svg}
\usepackage{booktabs}
\usepackage[font={small}]{caption}
\usepackage[group-separator={,}]{siunitx}

\DeclareMathOperator{\argmin}{arg\,min}
\graphicspath{{images/},{figures/}}

\title{End-to-end spoofing detection with raw waveform CLDNNs}
\name{Heinrich Dinkel, Nanxin Chen, Yanmin Qian, Kai Yu\thanks{This work was supported by the Shanghai Sailing Program No. 16YF1405300, the Program for Professor of Special Appointment (Eastern Scholar) at Shanghai Institutions of Higher Learning, the China NSFC projects (No. 61573241 and No. 61603252) and the Interdisciplinary Program (14JCZ03) of Shanghai Jiao Tong University in China}}
\address{Key Laboratory of Shanghai Education Commission for Intelligent Interaction and Cognitive Engineering\\
    SpeechLab, Department of Computer Science and Engineering \\
    Shanghai Jiao Tong University, Shanghai, China \\
    {\small \tt \{heinrich.dinkel,bobchennan\}@gmail.com,\{yanminqian,kai.yu\}@sjtu.edu.cn}
    }
\begin{document}
%\ninept
%
\maketitle

\begin{abstract}
Albeit recent progress in speaker verification generates powerful models, malicious attacks in the form of spoofed speech, are generally not coped with. Recent results in ASVSpoof2015 and BTAS2016 challenges indicate that spoof-aware features are a possible solution to this problem. Most successful methods in both challenges focus on spoof-aware features, rather than focusing on a powerful classifier. In this paper we present a novel raw waveform based deep model for spoofing detection, which jointly acts as a feature extractor and classifier, thus allowing it to directly classify speech signals. This approach can be considered as an end-to-end classifier, which removes the need for any pre- or post-processing on the data, making training and evaluation a streamlined process, consuming less time than other neural-network based approaches.
The experiments on the BTAS2016 dataset show that the system performance is significantly improved by the proposed raw waveform convolutional long short term neural network (CLDNN), from the previous best published 1.26\% half total error rate (HTER) to the current 0.82\% HTER. Moreover it shows that the proposed system also performs well under the unknown (RE-PH2-PH3,RE-LPPH2-PH3) conditions.
\end{abstract}
\begin{keywords}
CLDNN, End-to-End, BTAS2016, Spoofing detection
\end{keywords}
\section{Introduction}
\label{sec:intro}

Biometric recognition is a broad field which has developed from the classic fingerprint over to face recognition and nowadays speech can be used to naturally to restrict access to a certain medium. The research field which focuses on protecting the integrity of this speech based process is called speaker verification (SV). The main purpose of speaker verification is to detect whether the (real) speaker, who registered himself with the system, produces an utterance to grant access to the system or if that utterance was produced by an impostor. Malignant spoofing attacks mimic real speakers characteristics, thus an unprepared system's performance degrades heavily, when exposed to spoofing attacks \cite{Erguenay2015On,Evans2014Speaker}.
Traditional SV systems are not aware of possible spoofing attacks, which can be threatened by \textit{direct} attacks (also called \textit{spoofing} attacks). These attacks either artificially or naturally produce a spoofed utterance and try to gain access to a system. This work focuses only on the prevention of direct attacks. Overall there are currently four known \textit{direct} attacks (Impersonation, Replay, Synthesis, Voice conversion).
%
%Overall \textit{direct} spoofing attacks can be categorized into the following four types:
%\begin{itemize}
%\item \emph{Impersonation}: A (generally adept) speaker mimics another person. These attacks gained little interest in recent research and are generally seen as being less potent, whereby to be effective, this approach requires professional expertise. Skilled impersonators, although being a potentially serious threat, are a hazard that generally does not occur often.
%\item \emph{Replay}: An already recorded utterance of a speaker is played into the SV system. Nowadays smartphones are widely available and can easily be used to record and replay speech, which makes this approach an increasingly practical one. For replay attacks, replay aware features, such as RPS \cite{Villalba2015} and channel noise detection \cite{Wang2011} seem to be effective countermeasures.
%\item \emph{Synthesis (TTS)}: A speech synthesis system is used to generate spoofed utterances from written text. Synthesis becomes an increasingly big threat to SV systems due to the increasingly large availability given by open source libraries. In addition, intensive research towards synthesis systems makes them a serious threat for SV systems.
%\item \emph{Voice conversion}: Similar to speech synthesis this attack type focuses on converting prerecorded speech from any person in order to mimic specific vocal features of a target speaker. This approach is likely to be as dangerous as TTS, successful countermeasures to detect voice conversion incorporate LPC and LPCC features \cite{Alegre2013a}.
%\end{itemize}

Building an appropriate feature representation and designing a suitable classifier for each of the attack types are seen as separate problems, with different approaches for a suitable solution. One of the general disadvantages is that these features might not be optimal for the classification succeeding the classification task. Our motivation stems from recent advances in anti-spoofing research, which shows that an appropriate feature - independent of the classifier - contributes to prevent spoof attempts of a speaker verification system. In this context, deep neural networks can be seen as a joint classification and feature extraction framework, that aim to generate a feature representation which incorporates all relevant \textit{direct} attack types.

%While traditional features are preprocessed using filter-banks, which are not adjusted to the task at hand, raw wave input can be used to construct a model which is capable of learning the filters and filter-banks jointly with the rest of the network. This model can then be used to extract deep features, which when extracted from lower layers of a deep neural network (DNN) represent the speaker information, while higher layers can be used for frame level representations \cite{Glorot2010}\cite{Liu2015}. %The LY 

%Within the classical DNN framework, the amount of information given by a single frame is not sufficient to let the network learn about larger chunks of frames (e.g. utterances), thus single frames are extended by a context window, that allows the DNN to fetch the rich information within a chunk.
%In order to allow this model to learn a context of larger chunks, a recurrent neural network with long short term memory (RNN-LSTM) is deployed to map a given input sequence of time-frequency filtered frames into a single valued representation. 

%
%In our experiments we picked two datasets namely BTAS2016 and ASVSpoof2015, which incorporate all of the shown above attack types. Precisely, the BTAS2016 dataset focuses mostly on replay speech attacks that bypass the microphone, compared to ASVSpoof2015 which corpus centers around synthesis attacks.

The remainder of this paper is organized as follows. At first \Cref{sec:prev} reviews previous work in the context of spoofing detection. Continuing with \Cref{sec:cldnn}, which describes the CLDNN architecture with raw wave input. \Cref{sec:experiment} describes our experimental setup, model parameters, used datasets and demonstrates the results while comparing the CLDNN approach with other neural network anti-spoof techniques. A conclusion is given in \Cref{sec:conclusion}.

\section{Previous work}
\label{sec:prev}

%Most Neural networks in speaker verification and anti spoofing tasks are trained using conventional MFCC or PLP features. Both features make use of different filterbanks. These filter banks are fixed-sized ( e.g. mel filter bank, bark filter bank), thus can be suboptimal for the task they are used for. %The MFCC and PLP features have a fixed sized filter bank in common. It can be assumed that this filterbank is inferior to a filterbank which could be learned by a neural network.

\subsection{Model}
\label{sec:model}
Previous models include the ever so popular {\it i}-vector \cite{Weng2015Sysu,Novoselov2016Stc,Khoury2014Introducing}, which when used as a standalone model does not perform well. Another popular approach is to use the traditional GMM model. In this approach two different GMMs are trained representing the \textit{genuine} ($M_g$) and \textit{spoof} ($M_s$) labels respectively. Each GMM only uses the respective training data. After training, the score for a given evaluation utterance $\mathbf{x}$ can be calculated as follows:
\begin{equation}
\text{score}(\mathbf{x}) = \log P(\mathbf{x} | M_g ) - \log P (\mathbf{x}| M_s )
\end{equation}
Here $P(\mathbf{x})$ is assumed to be Gaussian distributed. Successful attempts can be seen in \cite{Todisco2016New,Sahidullah2015Comparison}. Deep features were also employed and achieved remarkable results. Chen et al. \cite{Chen2015MultiTask,Chen2015Robust}, fed PLP features into a neural network and obtained a highdimensional representation vector. This vector is then used as basis for future classification. In recent work \cite{Qian2016Deep}, sequence models such as recurrent neural networks - long short term memory (RNN-LSTM) were incorporated to extract features. %Another powerful approach by Villalba et. al \cite{Villalba2015} used RPS Spectrum features and fed these into two separate neural nets. %In order to reduce the risk of over fitting, an independent one-class SVM was trained to detect abnormalities within data, which then was fused with the networks to achieve remarkable results.
%Compared to the traditional feature extraction process, which use mathematical models as their base, deep feature approaches use a neural network for feature extraction. 
%Deep features....
%Other deep feature approaches include \cite{}, \cite{}, where Bi-directional LSTM's are used to extract utterance-level features, which are then scored using LDA, Gaussian density function (GDF) or cosine distance.

\subsection{Features}

As previous works indicate, feature extraction is crucial in order to detect possibly malicious system accesses. Research towards detecting synthesized spoofing attempts shows that this type of speech generally generates artifacts that can be detected by features that use phase spectrum information. It was seen that phase spectrum based features (e.g. MGDF) \cite{Wu2012Detecting} seem to discriminate better than common magnitude ones. Moreover, the recently published constant Q cepstral coefficient (CQCC) feature can be seen as the current state-of-the-art, being capable to detect synthesis based attacks \cite{Todisco2016New}. Furthermore, deep feature approaches apply neural networks for feature extraction \cite{Liu2015Deep}. Deep feature frameworks use neural networks to achieve a high abstract representation of input frames \cite{Chen2015MultiTask}, in order to extract features from one of its hidden layers, which are then scored by using an independent classifier (e.g. GMM, SVM, LDA).

\section{Raw Waveform CLDNN Architecture}
\label{sec:cldnn}
The CLDNN architecture ( Convolutional LSTM Deep Neural Network ) combines three different types of neural networks into a single model. This model obtains an input in form of a sequence of frames and outputs a likelihood for the whole sequence. The CLDNN performs time-frequency convolution to reduce spectral variance, long-term temporal modeling by using a LSTM, and classification using a DNN.

\begin{figure}[!htb]
% \begin{minipage}[b]{1.0\linewidth}
\centering
\includegraphics[width = 5.5cm]{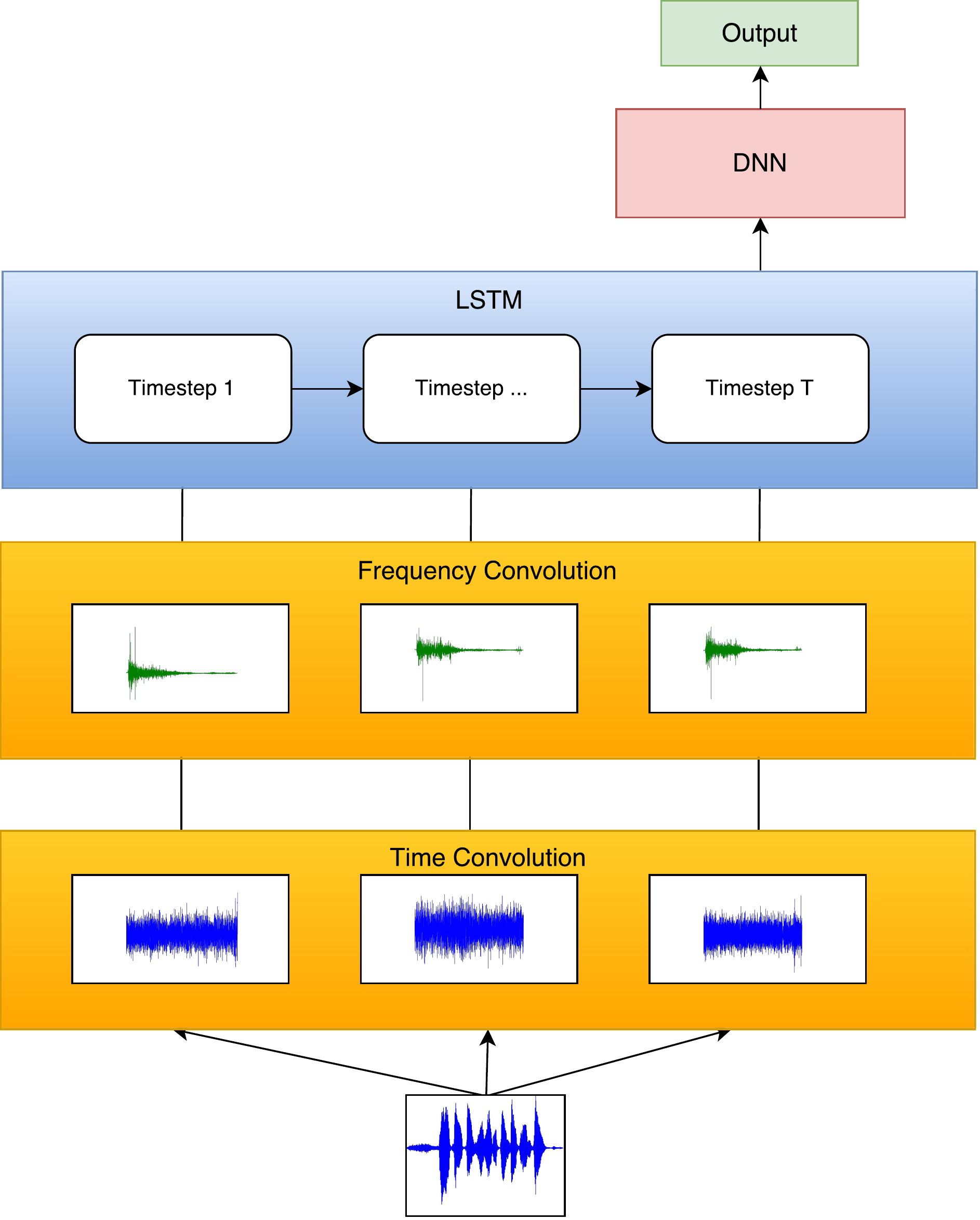}
  %\centerline{\includegraphics[width=8.8cm]{}}
%  \vspace{2.0cm}
  \caption{The architecture of the CLDNN with raw waveform}\label{fig:cldnn}\medskip
% \end{minipage}
\end{figure}

CLDNN was already successfully employed in automatic speech recognition tasks (ASR) \cite{Tobergte2013Convolutional}. In contrast to the more common approach to use log mel features, this approach uses raw waveforms as input. We argue that it might be more beneficial for the network to directly learn the time-frequency transformation process on top of being able to retain all information present within the time-domain, instead of having an already preprocessed, abstract, log-mel spectrum domain as input. Thus the network can learn dependencies between adjacent frames in the time domain and it's corresponding frequency domain transformation. In this work, a raw waveform based CLDNN is proposed for the spoofing detection, and the model architecture is specified as Figure \ref{fig:cldnn}. In this model the convolution is applied over a sequence of input features $\left[ x_1, \ldots, x_t ,\ldots, x_S \right] $. The convolutional layers are adjusted to share their parameters over the whole sequence with length $S$.

The first layer in this architecture is a time-convolutional layer over the raw time-domain waveform which can be thought of as a finite impulse-response filter bank followed by a non-linearity ( Rectified Linear Unit ) \cite{He2015Deep}. %The convolution uses a kernel to convolve the given samples into a higher dimensional representation. Assuming the input frame at time $t$ is denoted as $x_t$, thus the time-convolution layer uses some $P_{\text{time}}$ filters for convolution to represent $x_t$ with $P$ dimensions. Time convolution uses 160 samples ( 10 ms ) to shift the kernel and kernel-width is chosen to be equal to 400 (25 ms).
After time convolution, non overlapping max pooling is applied to remove any time variance, thus collapsing all of the input  samples ($N$) into a single value.  %Therefore, a single $P_{\text{time}}$ dimensional feature is obtained, further referred as $\tau_t(x)$.
Having collapsed the time-impulse to a smaller representation vector, frequency convolution follows the time convolution to reduce the phase variations within the time-filtered signal. %The frequency layer uses a similar setup as in \cite{Sainath}, using $F_{\text{freq}}$ filters with a kernel-width of $8$. Non overlap max pooling is then applied where a kernel width of $3$ is used. 
After frequency convolution, the output for each time-step $t$ is then fed into a two layer LSTM, which outputs a sequence of fixed sized vector representations. In this paper, the last time-step of the sequence is picked to obtain a single vector representation, which then is fed into the neural network classifier, as shown in \Cref{fig:cldnn}. 
DNN, CNN and the LSTM are jointly trained within the network. The last layer of the DNN utilizes softmax activation that normalizes the outputs to sum to one.
During training a sequence of samples of length $S$ is taken from each utterance and fed it into the network. Evaluation is performed by inputting a whole utterance at a time into the network which produces likelihood scores at the output layer for each class. In this task, four output neurons (each for one attack type (SS, VC, RE) plus the genuine speakers) are used. Thus, the final score is obtained by taking the log-likelihood values which correspond to the genuine class as scores. Thus larger scores correspond to genuine speakers, lower scores correspond to spoofed speech.

% \begin{figure}[htb]
% \begin{minipage}[b]{1.0\linewidth}
%   \centerline{\includegraphics[width=3.9cm]{Overview_CLDNN.png}}
% %  \vspace{2.0cm}
%   \caption{Basic CLDNN structure}\label{fig:reddot_err}\medskip
% \end{minipage}
% \end{figure}
% \begin{equation}
% Score(X_eval) = P( X_eval | C = genuine ) 
% \end{equation}
\subsection{Normalization}

The input of the network is directly taken from the raw waveform of a given audio utterance. One utterance with $L$ samples will be cut into same sized pieces of length $N$. Pieces with length $<N$ will be removed. It is important to note that no voice activity detection is utilized, because artificially created speech sometimes has unusually long silenced segments, which when removed, can degrade the model's performance. In this work, mean and standard deviation normalization are applied onto the extracted raw waves, resulting in unit mean and zero variance within the input data.

\subsection{Model specification}

The front-end of our framework is comprised of two convolutional layers that aim to invariantly transform the signal in the time and frequency domain \cite{Hoshen2015Speech}. After each convolution batch normalization is used and followed by a non-overlapping maxpooling. Throughout the network, rectified linear unit (ReLU) is used as activation function. 
The model is further extended by dropout \cite{Srivastava2014Dropout} a probability of $50\%$, between each linear layer in the classifier, as well as after each LSTM layer. The number of time-kernel convolution filters is set as $39$ ( comparable to MFCC/PLP ), the sequence length $S$ describes the number of following frames which are fed into the CLDNN. 

%It was discovered that networks with randomly initialized weights perform at least as well as pretrained ones \cite{He2015a}. Therefore, research focused on discovering new weight initialization techniques in order to increase the learning rate of a neural network \cite{Glorot2010}\cite{Sutskever2012}.
As optimization method it was decided to use an adaptive learning algorithm, adadelta \cite{Zeiler2012Adadelta}. As described in \cite{Sainath2015Learning}, using a larger frame size $N = 560$ while keeping the kernel width at $400$ is beneficial for the final performance. Moreover we use a stride of $160$ (10 ms) as step for the time-convolution filter.
% In all of our experiments the number of time-filters $P$ is set to be $39$ (\Cref{tab:cldnn_model}).

\section{Experiments}
\label{sec:experiment}

% \subsection{Deep feature classifiers}

% In this paper our research focuses on using Softmax and GDF as our primary classifiers.
% \subsubsection{SoftMax}

% {\bf Softmax} labeled result use the output-probability of the neural network as a score. After training the model, the evaluation utterances are fed into the network, which uses its last ( Softmax ) layer to obtain the respective class probabilities. Advantage of this method is that, besides forward pass computation, no other calculation is required. 

% \subsubsection{GDF}
% {\bf GDF} (Gaussian density function) is a classical classifier following a generative manner, speaker identity vector representations(e.g. {\it i}-vector) are modeled by a Gaussian distribution, where full covariance matrix is shared across all speakers. For an test vector representation $\mathbf{w}$, we evaluate the log likelihood score against the target $m$,

% \begin{equation}
% \log(\mathbf{w}|m)=\mathbf{w}^T\mathbf{\Sigma}^{-1}\mathbf{\mu}_m-\frac{1}{2}(\mathbf{w}^T\mathbf{\Sigma}^{-1}\mathbf{w} +\mathbf{\mu}_m^T\mathbf{\Sigma}^{-1}\mathbf{\mu}_m) 
% \end{equation}

% where $\mathbf{\mu}_m$ is the mean vector for speaker $m$, $\mathbf{\Sigma}$ is the common covariance matrix and $const$ is a speaker- and vector representation-independent constant. Furthermore, the speaker-independent part can be neglected, leading to:
% \begin{equation}
% \log(\mathbf{w}|m)=\mathbf{w}^T\mathbf{\Sigma}^{-1}\mathbf{\mu}_m-\frac{1}{2}\mathbf{\mu}_m^T\mathbf{\Sigma}^{-1}\mathbf{\mu}_m
% \end{equation}

\subsection{Dataset}

In this paper, the focus lies on the BTAS2016 dataset \cite{Korshunov2016Overview}, having overall \num[group-separator={,}]{43553} utterances of training, \num[group-separator={,}]{43575} utterances of development and \num[group-separator={,}]{50496} utterances of evaluation data. The emphasis of this dataset lies on replay attacks, which include low and high quality laptop as well as phone recordings. Unknown replay attacks were also recorded on laptop and phone devices, but differ from these used in the training set.

%
%\begin{table}[!tbh]
%\centeringincludegraphics
%\small
%\caption{Dataset structure. TTS is text-based synthesized speech, VC is voice conversion. Replay attacks are structured into known (are seen during training) and unknown (only in testset)}
%\label{tab:database}
%\begin{tabular}{|l||ccc|}
%\hline
%{\bf Type of data} & {\bf Train }& {\bf Dev } & {\bf Test } \\ 
%\hline
%\hline
%genuine data & 4973 & 4995 & 5576 \\ 
%\hline
%\hline
%all attacks & 38580  & 38580 & 44920\\ 
%\hline
%\hline
%TTS & 980 & 980 & 1120 \\ \hline
%VC & 34800 & 34800 & 39000 \\ \hline
%Known Replay & 2800 & 2800 & 3200 \\ \hline
%{\bf Unknown Replay} & - & - & 1600\\ \hline
%\end{tabular}
%\end{table}

%Shown in \Cref{tab:database}, known replay attacks include low and high quality laptop as well as phone recordings. Unknown replay attacks were also recorded on laptop and phone devices, but differ from these used in the training set.

\subsection{Evaluation Protocol}

First, a model is trained on the provided training data. After training, the development utterances are inserted into the system and scores for each utterance are obtained, which are consequently used to compute the FAR and FRR.
The \textit{false acceptance rate} (FAR) and the \textit{false rejection rate} (FRR) are both metrics, which depend on a certain threshold $\theta$:

\begin{align}
\text{FAR}(\theta) &= \frac{\vert\text{score}_{\text{attack}} \geq \theta \vert}{\vert\text{score}_{\text{attack}}\vert},
\text{FRR}(\theta) &= \frac{\vert\text{score}_{\text{real}} < \theta \vert}{\vert\text{score}_{\text{real}}\vert}
\end{align}

As evaluation metric, the \textit{half total error rate} (HTER) \cite{Chingovska2014Biometrics} is used. The threshold of the development data $\theta_{dev}$ is utilized, in order to calculate the FRR and FAR of the evaluation data:
\begin{align}
\theta_{dev} &= \argmin_{\theta} \left( \frac{\text{FAR}_{\text{dev}}(\theta) + \text{FRR}_{\text{dev}}(\theta)}{2}\right)\\ 
\text{HTER}_{\text{eval}} &= \frac{\text{FAR}_{\text{eval}}(\theta_{\text{dev}}) + \text{FRR}_{\text{eval}}(\theta_{\text{dev}})}{2}
\end{align}
%Scoring the subcategories is likewise done by using the $\text{FAR}_{\text{dev}},\text{FRR}_{\text{dev}}$ thresholds.

\subsection{Baseline}
\label{sec:baseline}

The baseline uses a standard GMM approach, which is trained using the procedure defined in \Cref{sec:model}. In traditional SV tasks, feature extraction is assisted by voice activity detection (VAD). In light of spoofing detection, VAD is not applied, since SS and VC methods tend to create unnaturally long silent segments. Thus, silence is a key factor in determinating artificially created speech (SS,VC categories). 
%

% \begin{table}[!htb]
% \centering
% \small
% \begin{tabular}{| p{10mm} | p{10mm} | p{12mm} | p{12mm} | p{8mm} |}

% \hline
% Window \ size & Window \ shift & Static \ dimension & Normal- \ ization & Dynamic \ dimension \\
% \hline
% \hline
% 25ms & 10ms & 13 & Mean+Var & 39 \\

% \hline
% \end{tabular}
% \caption{PLP parameters}\label{tab:plpparams}
% \end{table}

\begin{table}[!htb]
\centering
\small
\begin{tabular}{| c || c |}

\hline
Parameter & Value  \\
\hline
\hline
Windowsize & 25 ms \\
\hline
Windowshift & 10 ms \\
\hline
Static dimension & 13 (12 ceps + power) \\
\hline
Normalization & Cepstral mean + var\\
\hline
Dynamic dimension & static + $\Delta + \Delta \Delta$ = 39  \\

\hline
\end{tabular}
\caption{PLP parameters}\label{tab:plpparams}
\end{table}

The GMM baseline uses 512 Gaussian components. The training procedure is the same as described in \Cref{sec:model} and uses the features \Cref{tab:plpparams}. The baseline GMM and the formal published BTAS2016 challenge results are shown in \Cref{tab:baseline}.

\begin{table}[!htb]
\centering
\small
  \begin{tabular}{|c|c|c||c|}
  \hline
   Placing & Model   & Feature      &   HTER \\
   \hline
   \hline
   Baseline & GMM & PLP-39       &   2.96 \\
   \hline
   3rd & BLSTM-DNN & PLP-39     &   2.20 \\
   2nd & LDA     & Spectral-M-V &   2.04 \\
   
   1st & GMM & MFCC+i-MFCC  &   \bf{1.26} \\
  \hline
  \end{tabular}
  \caption{Previous results for BTAS2016 \cite{Korshunov2016Overview}}\label{tab:baseline}
\end{table}

%From \Cref{tab:baseline}, it can evidently be seen that the GMM-UBM classifier performance depends heavily on the front end features. %Consequently, powerful features play a more significant role in determinating the final performance compared to the classifier itself. 

% Baseline results (\cref{tab:baseline}) indicate that a direct classification with PLP features is not sufficient. 

% \begin{table}
% \centering
% \begin{tabular}{| c || c | c || c |}

% \hline
% \multirow{2}{*}{$N$(framewindow)} & $W$& init & EER \\ 
% %  & $W$(time width) & $P$(time) & $F$(freq) & init & EER\\
%  \hline 
%  \hline
%  400 (25 ms) & 400 (25 ms) & random & 0 \\
%  \hline 
%  400 (25 ms) & 400 (25 ms) & kaiming & 0 \\
%  \hline
%  560 (35 ms) & 400 (25 ms) & kaiming & 0 \\
%  \hline
%  700 (42 ms) & 400 (25 ms) & kaiming & 0 \\
%  \hline
% \end{tabular}
% \caption{Framesize analysis (Softmax)}\label{tab:framesize}
% \end{table}

%\section{Results}
\subsection{CLDNN - Setup}
\label{sec:results}

In our experiments we see that the features extracted from the front-end CNN are generally rich enough in information content. We adapt the same architecture as seen in \cite{Sainath2015Learning}, but do tune our model to fit the dataset better. Two different setups are presented in \Cref{tab:cldnn_model}, a large CLDNN-1 model, which acts as the basemodel and a smaller CLDNN-2. %We use the smaller CLDNN-2 (\Cref{tab:cldnn_model}) to verify if the large CLDNN-1 might overfit the data. %After running CLDNN-1 with different configurations, we discovered that an increase in model complexity does not commensurate with an enhancement of the final EER. Thus a smaller model (CLDNN-2 as in \Cref{tab:cldnn_model}) was created, to test if CLDNN-1 overfits the data.

\begin{table}[!htb]
\centering
\small
\begin{tabular}{| c || c | c | c | }

\hline
Setup & CNN-Maps & LSTM & DNN \\
\hline
\hline
CLDNN-1 & 39 (time) + 256(frequency) & 256 & 512 
\\ \hline
CLDNN-2 & 39 (time) + 128(frequency) & 128 & 256\\
\hline
\end{tabular}
\caption{CLDNN Setup. All of the models use a two convolutional layers (time + frequency), two layers of LSTM and a single layer of DNN.}\label{tab:cldnn_model}
\end{table}

\subsection{Sequence length influence}
\label{sec:seq}

The sequence length plays a crucial role in training RNN-based systems, thus the question arises if the CLDNN performance increase commensurates with a larger sequence length, as it is the case for standard LSTM. It is investigated which sequence length is most likely to be optimal for this task.  %In this experiment we use CLDNN-1, with the same parameters as described in \cref{sec:model}, having a framesize of $560$ and a kernel width of $400$.

\begin{table}[!htb]
\centering
\small
\begin{tabular} {| c | c | c ||  c |}
\hline
Sequence length & FAR & FRR & HTER \\
\hline
\hline
25 & 2.98 & 0.14 &  1.56 \\
\hline
50 &  2.62 & 0.79 &  1.7 \\
\hline
70 &  3.56 &  0.8 &  2.18\\
\hline
\end{tabular}
\caption{Sequence length influence, note that FAR and FFR are taken from HTER Error. CLDNN-1 is used as model.}\label{tab:frameinfluence}
\end{table}

The results using multiple sequence lengths show an uncharacteristic behavior for RNN's. We assume that either the CNN front-end contribution to the final performance is more significant than the LSTM or that by increasing the sequence-length inadvertently decreases the number of samples available in the dataset, which leads to a possible underfit of the data. 

%The results of this comparison show that the model can classify short term information easier. Moreover it can be seen that the CLDNN-1 results outperform the PLP-39 baseline in \Cref{tab:baseline}, thus further investigation if the current best result on the dataset can be beaten by the CLDNN model is done. 

\subsection{Neural network comparison}

\begin{table}[!htb]
% \scriptsize
% \footnotesize

\small
\label{tab:details}
\begin{tabular}{|l|p{5mm}|p{6mm}|p{6mm}|p{6mm}|p{6mm}|p{7mm}|}
\hline
{\bf Attack} & {\bf LS TM} & {\bf BL STM}  & {\bf BL STM DNN} & {\bf CLD NN-1} & {\bf CLD NN-2} & {\bf Best-BTAS} \\ 
\textit{Classifier} &  LDA & LDA & LDA & Soft max & Soft max & GMM \\
\hline
\hline
All &     		   	 2.99		&    2.43 & 		 1.21		    & 	     1.56 &  \textbf{0.82}  & \textit{1.26}  \\  \hline
\hline
SS-LP-LP &         	 1.51		&    1.79 & 		 0.5		    &        1.14 &    \textbf{0.38}  & 0.68 \\ \hline
SS-LP-HQ-LP &      	 1.42		&    1.61 & 		 0.95		    &        1.05 &    \textbf{0.64}  & 0.68 \\ \hline
\hline
VC-LP-LP &         	 2.64		&    1.89 & 		 0.59		    &        0.97 &    \textbf{0.49}  & 0.68 \\ \hline
VC-LP-HQ-LP &      	 1.64		&    1.9  & 		 0.57		    &        0.55 &    \textbf{0.33}  & 0.81 \\ \hline
\hline
RE-LP-LP &         	 4.1		&    2.85 & 		 1.06		    &         0.63 &    \textbf{0.52}  & 0.87 \\ \hline
RE-LP-HQ-LP &      	11.54		&    8.54 & 		 6.69		    &         2.88 &    \textbf{0.96}  & 1.81 \\ \hline
RE-PH1-LP &        	 5.73		&    2.04 & 		 0.69		    &         0.57 &    \textbf{0.52}  & 0.68 \\ \hline
RE-PH2-LP &             3.29     &   1.54 & 		 \textbf{0.5}		    &         0.57 &    1.08  & 0.68 \\ \hline
\hline
{\bf RE-PH2-PH3}   &    9.48   	&    4.29 &      3.06      		    &    	 6.63 &	 \textbf{1.33}  & 6.49 \\ \hline
{\bf RE-LPPH2-PH3} &   27.35 	&   22.1  &     26.44      			&     	38.07 &   \textbf{21.14} & 23.06 \\ \hline

\end{tabular}
\caption{Comparison between other NN-approaches and the currently best result. Note that "RE-LP-LP","VC-LP-LP" and "SS-LP-LP" do not incorporate the HQ categories (in contrast to the original paper). }\label{tab:results}
\end{table}
For a better comparison to other neural network based methods, a two (LSTM), a two layer bidirectional LSTM (BLSTM) and an improved DNN-BLSTM fusion model, similar to that in \cite{Korshunov2016Overview} were also trained. LSTM and BLSTM models contain in each layer $512$ neurons. The DNN-BLSTM model uses the concatenated output vectors of a 7 layer DNN (from the 3rd hidden layer) in addition to the output of three different BLSTM models. The BLSTM output vectors have a size of 512 each, while the DNN uses $1024$ dimensional vector representations. Thus, a $1024+(512 \times 3) = 2560$ dimensional vector representation is obtained. Compared to the proposed end-to-end model, these models make use of a backend LDA classifier, which creates a single score for each vector representation. LSTM, BLSTM and BLSTM-DNN models all use a sequence length of $50$.The LSTM, BLSTM and DNN-BLSTM models uniformly use PLP-39 features, similar to these in \Cref{tab:plpparams} as their input. The results of this paper are compared with other neural networks attempts for antispoof (\Cref{tab:results}). Note that in the competitions paper the categories "LP-LP" were used as a superset of "LP-HQ-LP", thus being non independent, where in this work we assume each category is independent from each other. 
%The attack type abbreviation "SS" represents synthesized samples, "VC" stands for voice conversion and "RE" for replay attack. "LP" and "PH" indicate whether a phone or a laptop was used. "LP-RE" means that an utterance was recorded using a laptop and played to a phone. Note that in the competitions paper the categories "LP-LP" were used as a superset of "LP-HQ-LP", thus being non independent. In our results, we specifically separate all different result types. Therefore, the attacktype "RE-LP-LP" differs the most between our results and the original paper's. This separation does only affect the non "HQ" labeled results, nor the final result.

%\begin{figure}[!htbp]
%\centerline{\includegraphics[width=8.8cm]{figures/det_curves.pdf}}
%\caption{Det plot of evaluation results for those in \Cref{tab:results}}\label{fig:det}
%\end{figure}

%CLDNN-1 and CLDNN-2 both use a sequence length of $25$, and the configuration described in \Cref{tab:cldnn_model}. Network weights are uniformly initialized with the method \cite{He2015a}.

As it can be seen in \Cref{tab:results}, the BLSTM-DNN fusion model does outperform the baseline shown in \Cref{sec:baseline}, as well as other neural network based approaches. Additionally, it sets the mark of creating the currently best result on this corpus. Furthermore, the CLDNN-2 model performs well on unknown attacks (11.64\% compared to 14.78\%).  

%\begin{figure}[!htb]
%\centerline{\includegraphics[width=9cm]{eer_analysis_known_unknown.png}}
%\caption{Attack type error rates, categorized into known (SS + VC + RE) and unknown (RE-PH2-PH3 + RE-LPPH2-PH3) attacks}\label{fig:attacks}
%\end{figure}
% \begin{table}[!htbp]
% \centering
%   \begin{tabular}{| c | c || c | c | c | c |}
%     \hline
%   Model & Classifier  & $S$ &  FAR & FRR & HTER \\
% %  \cline{3-6}
% %  & & GC & PLDA \\
%   \hline
%   \hline
%   LSTM & Softmax & 50 & 0.0 \\
%   \hline
%   BLSTM & Softmax & 50 & \\
%   \hline
%   DNN-BLSTM\cite{} & LDA & 50 & 2.03 \\
%   \hline
%   CLDNN & Softmax & 25 & 0.0  \\
%   \hline
%   SS-LP-LP & & & 
%   \end{tabular}
 
% \end{table}

\section{Conclusion}
\label{sec:conclusion}

This paper successfully introduces an end-to-end framework using a raw waveform based CLDNN model for spoofing detection. Compared to the previous deep feature based methods, which builds the front-end and back-end separately, the new end-to-end raw waveform CLDNN makes the whole detection process more flexible, while at the same time being able to use the rich speech information from raw waveform by simultaneously optimizing feature extraction and classification accuracy. Surprisingly, performance increases by using these unprocessed raw waveform, indicating that raw signal might be a valid start point for this task, and the joint optimization on both front-end and back-end also makes this new architecture advanced. %Compared to conventional deep feature models, CLDNN does not rely on a backend classifier, thus making this approach more flexible. The proposed raw waveform CLDNN model is powerful when it comes to detect replay based spoofing attacks.

In future research, we would like to focus on more complex front-end feature extraction using more suitable time and frequency filtering networks.
%The proposed CLDNN model for anti spoof tasks, even though being less . An advantage of this model compared to others it is scalability. 
\vfill
\pagebreak

\bibliographystyle{IEEEbib}
% \bibliography{CLDNN_Paper}
\bibliography{Remote}

\end{document}